\documentclass[twocolumn, english, prl, floatfix, superscriptaddress, showpacs]{revtex4-1}

\usepackage[T1]{fontenc} 
\usepackage[latin9]{inputenc}
\usepackage{amstext} 
\usepackage{graphicx} 
\usepackage{physics}
\usepackage[retainorgcmds]{IEEEtrantools}

\usepackage{subfig} 
\usepackage{babel}

\begin{document}

\title{Irreducible many-body correlations in topologically ordered
  systems}

\author{Yang Liu} %
\affiliation{Beijing National Laboratory for Condensed Matter Physics,
  and Institute of Physics, Chinese Academy of Sciences, Beijing
  100190, China}


\author{Bei Zeng} %
\affiliation{Department of Mathematics \& Statistics, University of
  Guelph, Guelph, Ontario, Canada} %
\affiliation{Institute for Quantum Computing, University of Waterloo,
  Waterloo, Ontario, Canada}

\author{D.L. Zhou} %
\affiliation{Beijing National Laboratory for Condensed Matter Physics,
  and Institute of Physics, Chinese Academy of Sciences, Beijing
  100190, China}

\begin{abstract}
  Topologically ordered systems exhibit large-scale correlation in
  their ground states, which may be characterized by quantities such
  as topological entanglement entropy. We propose that the concept of
  irreducible many-body correlation, the correlation that cannot be
  implied by all local correlations, may also be used as a signature
  of topological order. In a topologically ordered system, we
  demonstrate that for a part of the system with holes, the reduced
  density matrix exhibits irreducible many-body correlation which
  becomes reducible when the holes are removed. The appearance of
  these irreducible correlations then represents a key feature of
  topological phase. We analyze the many-body correlation structures
  in the ground state of the toric code model in an external magnetic
  field, and show that the topological phase transition is signaled by
  the irreducible many-body correlations.
\end{abstract}


\maketitle


Topologically ordered phase may not be characterized by any local
order parameter associated with Laudau's symmetry breaking
picture~\cite{Wen2004}. How to characterize this type of exotic phase
is one of the
biggest challenges in modern condensed matter physics.
These topological phases may be characterized by many distinguished
features, including that: the degeneracy of ground states depends on
the topology of the manifold that supports the system; the existence
of anyonic elementary excitations; the existence of edge states on the
open boundaries. Moreover, all these properties characterizing
topological phase must be stable against local perturbations, making
topologically ordered systems promising candidates for fault-tolerant
quantum computing~\cite{Kit2003,NSS+2008}.

The topological entanglement entropy is firstly proposed to
characterize the ground state with topological order by
Kitaev-Preskill~\cite{KP2006} and Levin-Wen~\cite{LW2006}, which
builds a nontrivial connection between many-body physics and quantum
information. The underlying picture of the topological entanglement
entropy is to `retrieve' a many-body correlation which cannot be built
up from its parts. In general, calculating with large enough parts of
the system, the entanglement entropy is successfully used to identify
topological order in several microscopic
models~\cite{JWB2012,IHM2011,DMS2012}.

In this letter, we provide a novel perspective to signal this
many-body correlation in topologically ordered ground states, building
on the concept of irreducible many-body
correlations~\cite{LPW2002,Zho2008}. This approach intuitively sounds
as irreducible $r$-body correlation is nothing but the correlation
that cannot be build up from any $\leq (r-1)$-body
correlations~\cite{LPW2002,Zho2008}. In a topologically ordered
system, we demonstrate that for a region of the lattice with holes,
the reduced density operator exhibits irreducible many-body
correlations.




To be more precise, for an $n$-body quantum state and any $r\leq n$,
the irreducible $r$-party correlation (or irreducible correlation of
order $r$) characterizes how much information contained in the
$r$-particle reduced density matrices ($r$-RDMs) but not in the
$(r-1)$-RDMs. For a state $\sigma$, denote $C^{(r)}(\sigma)$ its
$r$-party irreducible correlation. The total correlation is then the
information contained in the state beyond that in the $1$-RDMs. In
this sense the irreducible $r$-party correlations provide a natural
hierarchy of correlations in the system -- the sum of all the
irreducible $r$-party correlations equals the total
correlation~\cite{Zho2008}.

Throughout the paper we consider lattice spin models. For any regions
$\mathcal{A},\mathcal{B}$ of the lattice, with
$\mathcal{B}\subset\mathcal{A}$, one naturally expects that there are
more correlations in $\mathcal{A}$ than those in $\mathcal{B}$, as all
the particles in $\mathcal{A}$ are contained in $\mathcal{B}$. This is
in general also the case for irreducible correlations. However
counter-intuitively, in topologically ordered systems, one could have
\begin{equation}
  \label{eq:redcor} 
  C^{(r)}(\rho^{\mathcal{A}}) <C^{(r)}(\rho^{\mathcal{B}}),
\end{equation}
where $\rho^{\mathcal{A}}$ ($\rho^{\mathcal{B}}$) is the reduced state
of the region $\mathcal{A}$ ($\mathcal{B}$). The extreme case could be
that $C^{(r)}(\rho^{\mathcal{A}})=0$ but
$C^{(r)}(\rho^{\mathcal{B}})>0$. This may happen when $\mathcal{B}$
and $\mathcal{A}$ have different topology (e. g.
$\mathcal{A}\setminus\mathcal{B}$ is a hole). In this case the
$r$-party correlation in $\rho^{\mathcal{B}}$ must become reducible, i.e. the information in the
$r$-RDMs of the region ${\mathcal{B}}$ is contained in the information
of the $r'$-RDMs of the region ${\mathcal{A}\setminus\mathcal{B}}$,
with $r'<r$.

It turns out that the appearance of these irreducible many-body
correlations in a region with holes, or the validity of
Eq.\eqref{eq:redcor}, represents a key feature of topologically
ordered systems. We will analyze these irreducible many-body
correlation structures in the ground state of the toric code model in
an external magnetic field. In addition to demonstrate the appearance
of these irreducible correlations and the nontrivial phenomena of
Eq.\eqref{eq:redcor} in these systems, we  show that the
topological phase transition is also identified by the creation of
these irreducible many-body correlations.


\textit{The Toric code mode}-- We start with the toric code
model~\cite{Kit2003}, which is a spin-$\frac{1}{2}$ model on an
$L\times L$ square lattice, with every edge representing a spin, hence
there are total $n=2L^2$ spins. The Hamiltonian is
\begin{equation}
  H_{\text{tor}}=-\sum_{s}A_{s}-\sum_{p}B_{p},
\end{equation}
where $s$ runs over all vertices (stars) and $p$ runs over all faces
(plaquettes). The star operator
$A_{s}=\prod_{j\in\partial{s}}\sigma_{j}^{x}$, where $\partial{s}$ is
the set of edges surrounding the vertex $s$. The plaquette operator
$B_{p}=\prod_{j\in\partial p}\sigma_{j}^{z}$, where $\partial p$ is
the set of edges surrounding the face $p$. In the above formula, the
operators $\sigma_{j}^{x}$ and $\sigma_{j}^{z}$ are Pauli operators of
the $j$-th spin.
\begin{figure}[h!]
  \label{fig:tormod}
  \centering
  \includegraphics{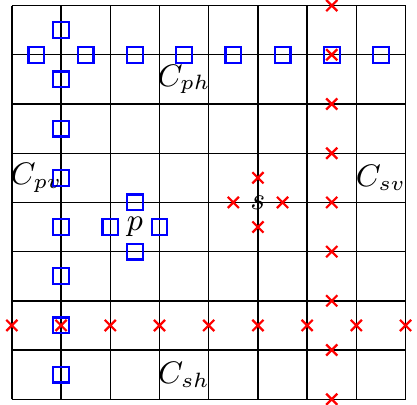}
  \caption{The toric code model. A blue rectangle denotes
    $\sigma^{z}$, and a red cross denotes $\sigma^{x}$. The plaquette
    operator $B_{p}$ and the star operator $A_{s}$ are demonstrated.
    The logic operators for the ground state space $\sigma_{I}^{x}$,
    $\sigma_{I}^{z}$, $\sigma_{II}^{x}$, $\sigma_{II}^{z}$ are
    associated with $C_{sh}$, $C_{pv}$, $C_{ph}$, $C_{pv}$
    respectively.}
\end{figure}

When the periodic boundary condition is considered (i.e., a torus),
the model shows typical features of topological order such as ground
state degeneracy. And the degenerate ground state space is a quantum
error-correcting code with macroscopic distance (i.e. the distance
grows with system size).  As the star and plaquette operators commute
the ground state space is given by the stabilizer formalism of quantum
code
\begin{equation}
  A_{s}\vert G\rangle = \vert G\rangle, \; B_{p}\vert G\rangle = \vert
  G \rangle. 
\end{equation} 

Furthermore, we have $\prod_{s}A_{s}=1$ and $\prod_{p}B_{p}=1$ for any
closed surface, which implies that there exist one dependent star
operator and one dependent plaquette operator.
The subspace of the ground states is characterized by the two sets of logical operators,
\begin{IEEEeqnarray}{rCl} \sigma_{I}^{x} = \prod_{j\in
    C_{sh}}\sigma_{j}^{x},&\;\;& \sigma_{I}^{z} = \prod_{j\in
    C_{pv}}\sigma_{j}^{z},\\ \sigma_{II}^{x} = \prod_{j\in
    C_{sv}}\sigma_{j}^{x}, && \sigma_{II}^{z} = \prod_{j\in
    C_{ph}}\sigma_{j}^{z},
\end{IEEEeqnarray}
hence the ground state degeneracy is $4$.


\textit{Irreducible many-body correlations in toric code model} -- We
start from considering the irreducible correlation for any ground
state $\ket{G}$ of the toric code model. Recall that for any $n$-qubit
state $\sigma$, the irreducible correlation
$C^{(r)}(\sigma)=S(\tilde{\sigma}^{(r-1)})-S(\tilde{\sigma}^{(r)})$,
where $\tilde{\sigma}^{(r)}$ is the $n$-qubit state with maximum
entropy, which has the same $r$-RDMs as those of $\sigma$, and $S$ is
the von Neumann entropy. And the total correlation
$C^{T}(\sigma)=\sum_{r=2}^n{C}^{(r)}(\sigma)=\sum_{i=1}^n
S(\sigma_i)-S(\sigma)$, where $\sigma_i$ is $1$-RDM of the $i$th
particle.

Since the $1$-RDMS of $\ket{G}$ are maximally mixed, the total
correlation of $\ket{G}$ is $C^{T}(\ket{G})=2L^2$. Furthermore, for
any $4\leq r\leq L$, the maximally mixed state $\rho_M$ supported on
the ground-state space has the maximum entropy among all states with
the same $r$-RDMs as those of $\ket{G}$. Therefore we have
\begin{eqnarray}
  \label{eq:1} 
  C^{(4)}(\ket{G}) & = & L^{2}-2,\\ 
  C^{(\ge L)}(\ket{G})& = &\sum_{i\ge L} C^{(i)}(\ket{G})
  = 2,
\end{eqnarray} 
and the irreducible correlations of all the other orders are $0$. This
implies that there are $2$ bits of irreducible correlations of
macroscopic order in any ground state of toric code model on a torus.



Notice that  $\rho_M$ itself contains and only contains irreducible
correlations of order $4$, in contrast to that $\ket{G}$ has
irreducible correlations of macroscopic order. Also the  thermal
state at temperature $T$ given by
$\rho(T)=\frac{e^{-\beta{H}}}{\textrm{Tr}e^{-\beta H}}$ (with
$\rho(0)=\rho_{M}$) also only contains irreducible correlations of
order $4$, according to the continuity argument~\cite{Zho2008}. In
general, for a system with $m$-body Hamiltonian, its thermal state
will contain irreducible correlations no more than order
$m$~\cite{CJZZ2012}.

If any of the ground state can show irreducible correlations higher
than  order $m$, then the ground states must be degenerate. And as
mentioned above, topologically ordered ground states exhibit
irreducible correlations of macroscopic order. This then raise an
interesting question:
where do these macroscopic correlations come from? To answer this
question, we examine in detail the correlations in the local reduced
states of the system.

\begin{figure}[htpb]
  \centering %
  \includegraphics[width=0.38\paperwidth]{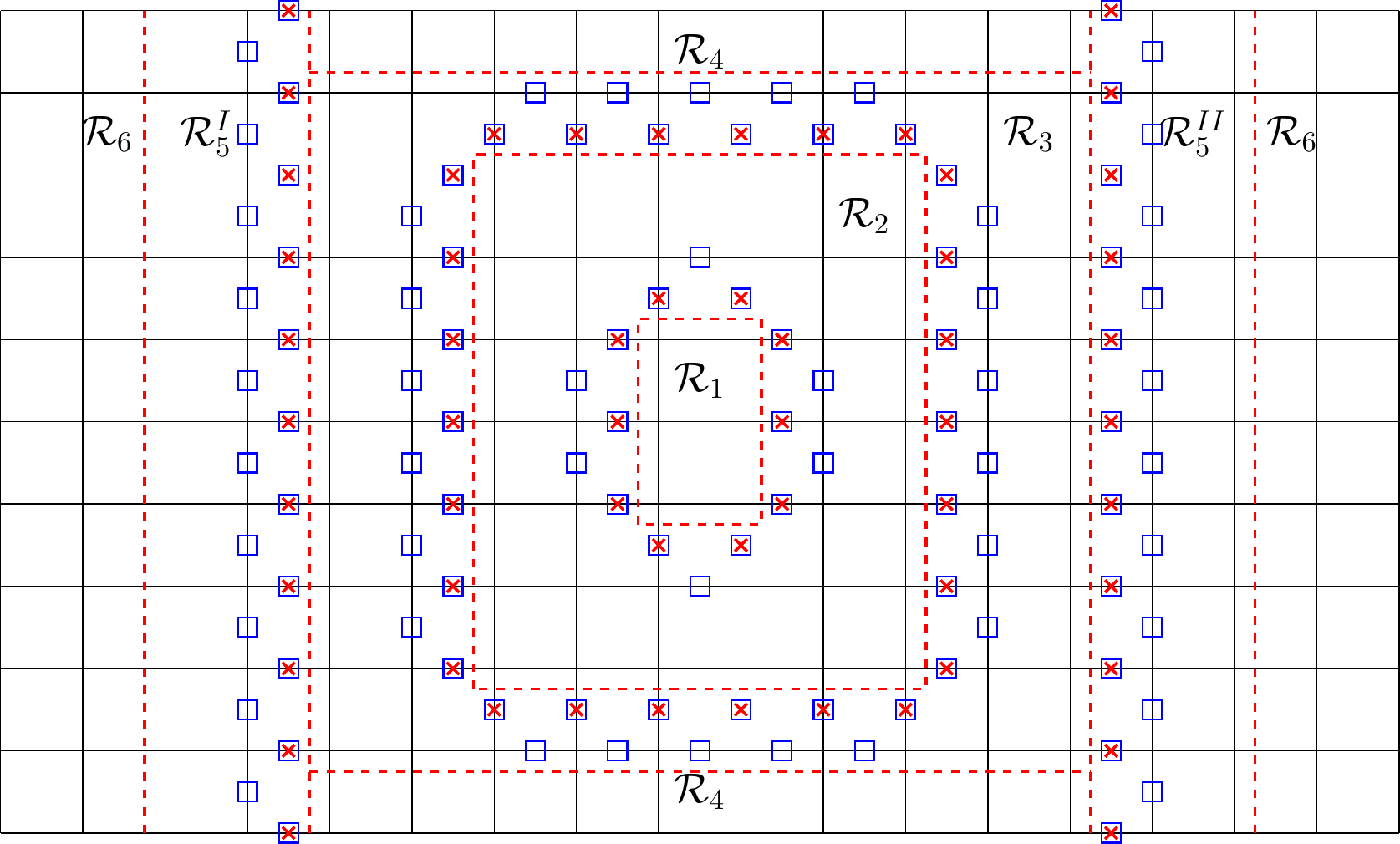}
  \caption{Reducible and irreducible multiparty correlations in the
    toric code model.The system is divided into seven parts, denoted
    as space $\mathcal{R}_{1}$, $\mathcal{R}_{2}$, $\mathcal{R}_{3}$,
    $\mathcal{R}_{4}$, $\mathcal{R}_{5}^{I}$, $\mathcal{R}_{5}^{II}$,
    and $\mathcal{R}_{6}$ by red dashed lines. Two irreducible
    multiparty correlations in $\mathcal{R}_{2}$ and $\mathcal{R}_{3}$
    are demonstrated, which also can be regarded as the reducible
    multiparty correlations in the joint region of
    $\mathcal{R}_{1}\cup \mathcal{R}_{2}\cup \mathcal{R}_{3}$. }
  \label{fig:cortor}
\end{figure}

Notice that when $r\leq L$, any $r$-RDM of $|G\rangle$ for a region of
the lattice without a hole is independent of the choice of the ground
states.
Furthermore, any such $r$-RDM has only irreducible correlations of
order $4$. As illustrated in Fig.\ref{fig:cortor}, this then implies
\begin{eqnarray}
  \label{eq:3} 
  C^{T}(\rho^{\mathcal{R}_{1}\cup \mathcal{R}_{2}\cup
    \mathcal{R}_{3}}) &=& C^{(4)}(\rho^{\mathcal{R}_{1}\cup
    \mathcal{R}_{2}\cup \mathcal{R}_{3}}),\\
  C^{T}(\rho^{\mathcal{R}_{1}\cup \mathcal{R}_{2}}) &=&
  C^{(4)}(\rho^{\mathcal{R}_{1}\cup \mathcal{R}_{2}}).
\end{eqnarray} 
Here we use dashed lines to divide the system into seven parts,
denoted as $\mathcal{R}_{1}$, $\mathcal{R}_{2}$, $\mathcal{R}_{3}$,
$\mathcal{R}_{4}$, $\mathcal{R}_{5}^{I}$, $\mathcal{R}_{5}^{II}$, and
$\mathcal{R}_{6}$ respectively.

However, if the region contains a hole, then situation could be
dramatically different. One can observe that
\begin{eqnarray}
  \label{eq:4} 
  && C^{(8)}(\rho^{\mathcal{R}_{2}})=C^{(14)}(\rho^{\mathcal{R}_{2}})=1,\\
  && C^{(26)}(\rho^{\mathcal{R}_{3}})=C^{(48)}(\rho^{\mathcal{R}_{3}})=1.
\end{eqnarray}
This implies that the reduced state in a region with a hole contains
$2$ bits irreducible correlations of orders proportional to the length
of the boundary. When the hole becomes larger and finally encounters
the boundary, the region splits into two parts, $\mathcal{R}_{5}^{I}$
and $\mathcal{R}_{5}^{II}$, and $2$ bits of irreducible macroscopic
correlations emerge.

Intuitively, these irreducible correlations in $\mathcal{R}_{2}$ are
contributed by the correlations in the reduced state of the hole
(${\mathcal{R}_{1}}$), which are reduced correlations in the region
with the hole ($\mathcal{R}_{1}\cup \mathcal{R}_{2}$). In general, for
any region $\mathcal{B}$ with holes, the reduced states exhibit
irreducible correlation of macroscopic order, which becomes reducible
for the region $\mathcal{A}$ without holes, where
$\mathcal{B}\subset\mathcal{A}$. And the existence of these
irreducible correlations of macroscopic order in regions with holes is
an essential feature of topological order.

\begin{figure}[htpb]
  \centering
  \includegraphics{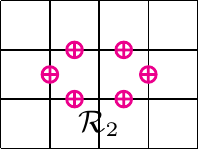}
  \caption{The region $\mathcal{R}_{2}$ contains $6$ spins labeled by
    magenta $\oplus$ on a $4\times{3}$ lattice. }
  \label{fig:conf}
\end{figure}

\textit{Characterizing topological phase transition} --
Based on the discussions above, it is natural to use the irreducible
correlations of orders proportional to the boundary length in a region
with a holes to signal the topologically ordered phase. As a typical
example, we consider the toric model in an external magnetic field
along the $\vec{n}$ direction \cite{DKO+2011}, with the Hamiltonian
\begin{equation}
  \label{eq:2} 
  H = H_{\text{tor}} - h \sum_{i} \vec{n} \vdot {\vec{\sigma}}_{i}.
\end{equation} 
Our calculation is based on the numerical exact diagonalization
method. Our system consists of $24$ spins on a $4\times{3}$ lattice,
and the irreducible $6$-particle correlation
$C^{(6)}(\rho^{\mathcal{R}_{2}})$ of $6$ spins in a region
$\mathcal{R}_{2}$ is studied, see Fig.~\ref{fig:conf}. Here it is
worthy noticing that $C^{(6)}(\rho^{\mathcal{R}_{2}})$ is not a
topological invariant, but it does reflect the power of creating
higher order correlations from lower order correlations.

In the thermodynamic limit, we expect that
$C^{(6)}(\rho^{\mathcal{R}_{2}})\neq{0}$ for a topological phase,
while it is zero for a non-topological phase. However, the numerical
value of $C^{(6)}$ might not be correct if calculating with finite
size systems. Nevertheless, even for the system of this small size,
the rate change of $C^{(6)}$ already clearly signals the phase
transition. In this sense we suggest to use the maximal changing rate
of $C^{(6)}$ with respect to $h$ as an indicator of the phase
transition point:
\begin{equation}
  \label{eq:5} 
  h^{*} = \mathop{\mathrm{argmax}}_{h}
  \dv{C^{(6)}(\rho^{\mathcal{R}_{2}}(h))}{h}.
\end{equation}

\begin{figure}[htpb]
  \centering \subfloat[]{ %
    \includegraphics{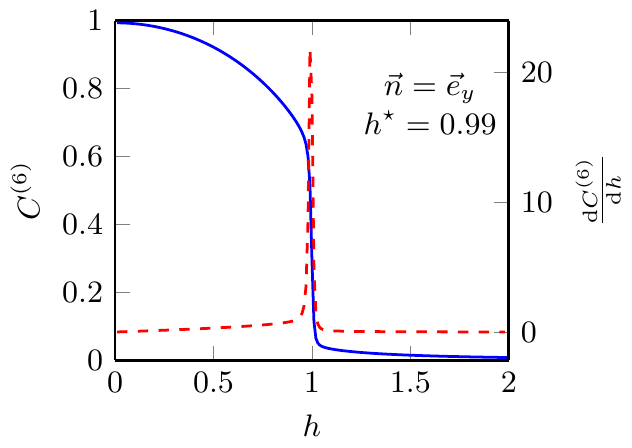}
    \label{a} }\\

  \subfloat[]{ %
    \includegraphics{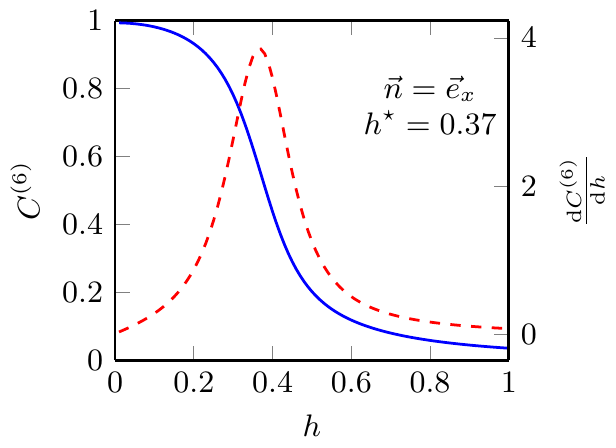}
    \label{b} }\\
  \caption{The $C^{(6)}(\rho^{\mathcal{R}_{2}})$ and its derivative
    varies with $h$ in the case of (a) $\vec{n}=\vec{e}_{y}$; (b)
    $\vec{n}=\vec{e}_{x}$.}
  \label{fig:sigtrans}
\end{figure}

Based on the numerical algorithms proposed in
Refs.~\cite{Zho2009b,NGKG2013}, we obtain the results of
$C^{(6)}(\rho^{\mathcal{R}_{2}})$ and its derivatives, for magnetic
field along both the $y$ direction and the $x$-direction, as shown in
Fig.~\ref{fig:sigtrans}. When the magnetic field is along the $y$
direction, there is a sharp transition behavior in
$C^{(6)}(\rho^{\mathcal{R}_{2}})$ with $h$, and the phase transition
point is $h^{\star}=0.99$, which is very close to the previous result
$h^{\star}=1$ in Refs.~\cite{VTSD2009,ODV2009}. When the magnetic
field is along the $x$ direction, $C^{(6)}(\rho^{\mathcal{R}_{2}})$
shows a smooth transition behavior, and the transition point is
$h^{\star}=0.37$, which is close to the previous result
$h^{\star}=0.34$ in Refs.~\cite{TKPS2010,WDP2012}.


\textit{General correlation structure in topological order} --
Although we discussed the toric code model, a similar correlation
structure should also be valid for topologically ordered systems in
general. In the thermodynamic limit, the ground states are degenerate
and any ground state exhibits the following correlation structure:
\begin{equation}
  \label{eq:6}
  C^{(r)}(\ket{G})=
  \begin{cases}
    \ge 0 & \text{if}\; r\le r_{0},\\
    = 0 & \text{if}\; r_{0}<r<L,\\
    \ge 0 & \text{if}\; r\ge L,
  \end{cases}
\end{equation}
where $r_{0}$ is a positive integer independent of the system size $L$. Furthermore, the value of irreducible correlations of
macroscopic order $C^{(\ge{L})}(\ket{G})$ is a topological invariant.

In a finite system, the ground state space is generally unique, which
does not exhibit any irreducible correlations of macroscopic order
i.e., $C^{(r)}(\ket{G})=0$ for $r>r_{0}$. However, there are always
irreducible correlations of higher order in the reduced states of
$\ket{G}$, which is manifested in a region with holes, e.g.
$\mathcal{R}_{3}$ in Fig.~\ref{fig:conf}, as the irreducible
correlations is of order proportional to the length of the inner
boundary $L_{I}$. Here we have
\begin{equation}
  \label{eq:7}
  C^{(r)}(\rho^{\mathcal{R}_{3}})=
  \begin{cases}
    \ge 0 & \text{if}\; r\le r_{0},\\
    \simeq 0 & \text{if}\; r_{0}<r<L_{I},\\
    \ge 0 & \text{if}\; r\ge L_{I}.
  \end{cases}
\end{equation}
When the region $\mathcal{R}_{3}$ is large enough (typically larger
than the correlation length), it then contains the irreducible
correlations
$C^{(\ge{L_{I}})}(\rho^{\mathcal{R}_{3}})=C^{(\ge{L})})(\ket{G})$, as
it is in the thermodynamic limit.
$C^{(\ge{L})})(\rho^{\mathcal{R}_{3}})$ therefore is the same
topological invariant as $C^{(\ge{L})}(\ket{G})$.

\begin{figure}[ht]
  \centering
  \includegraphics{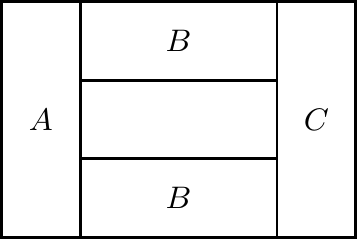}
  \caption{A region with a hole (e.g. $\mathcal{R}_{3}$) is divided
    into three parts $A,B,C$ in the scheme of Levin-Wen topological
    entropy. \label{fig:5} }
\end{figure}

\textit{The relation with topological entanglement entropy}-- The
construction of topological entanglement entropy by
Levin-Wen~\cite{LW2006}, denoted by $\mathcal{E}_{LW}$ may be regarded
as an approximation for obtaining the irreducible correlations of
macroscopic order in a large enough region with a hole, e.g. the
region $\mathcal{R}_{3}$. To calculate $\mathcal{E}_{LW}$, they
divided the region into three parts $A$, $B$, and $C$ as demonstrated
in Fig.~\ref{fig:5}, where $A$ and $C$ are far apart so they there
should be no correlation between them. It is worthy to point out that
here $A$, $B$, $C$ are taken as parties but not every single spin as before.

The Levin-Wen topological entropy is then given by the total
correlation
$C^{T}(\rho_{ABC})=S(\rho_{A})+S(\rho_{B})+S(\rho_{C})-S(\rho_{ABC})$
minus the biparty correlations $C^{T}(\rho_{AB})$ and
$C^{T}(\rho_{BC})$. For the correlation structure as shown in
Eq.~\eqref{eq:7}, it is reasonable to believe that only the
irreducible correlation $C^{(\ge{L_{I}})}(\rho^{\mathcal{R}_{3}})$
contributes to the three-party correlation among parties $A$, $B$, and
$C$. In fact, one can show that the irreducible three-party
correlation is upper bounded by such type of topological entropy, i.e.
$C^{(3)}(\rho_{ABC})\leq\mathcal{E}_{LW}(\rho_{ABC})$. The proof of
the inequality and the equality condition are given in the
supplementary material.

The topological entanglement entropy proposed by
Kitaev-Preskill~\cite{KP2006}, denoted by $\mathcal{E}_{KP}$, is
defined from the von Neumann entropy of a region without a hole, e.g.
the region $\mathcal{R}_{2}\cup \mathcal{R}_{1}$ in
Fig.~\ref{fig:conf}. The correlation between
$\mathcal{R}_{2}\cup\mathcal{R}_{1}$ and $\mathcal{R}_{3}\cup
\mathcal{R}_{4}$ comes from the dependence between $\mathcal{R}_{2}$
and $\mathcal{R}_{3}$ \cite{WVHC2008}, which implies the area law.
Notice that for the topological phase, the irreducible correlations of
macroscopic order in the region $\mathcal{R}_{2}$ will decrease the
correlations between $\mathcal{R}_{2}$ and $\mathcal{R}_{3}$, thus
decreases the entanglement entropy compared to the area law, which
then gives the topological entanglement entropy.

\textit{Summary} -- In sum, we use the concept of irreducible
many-body correlations to analyze the correlation structure in the
ground states of the toric code model in an external magnetic field.
The appearance of irreducible correlations of macroscopic order in a
region with holes represents an essential feature of topological
order. We also demonstrate that the power to create irreducible
correlations of higher orders for a region $\mathcal{B}$ with holes,
from irreducible correlations of lower order for a region
$\mathcal{A}\supset\mathcal{B}$, signals the topological transition
phase transition. Our calculation uses a relatively small system,
which clearly indicates the transition. Our concept has intimate
relations with the idea of topologically entanglement entropy and may
be applied to study other systems with topologically order, by
calculation with relatively small system size. Our work may shed light
on a better understanding of the general many-body correlation
structure of a quantum state in topologically ordered phase.

\begin{acknowledgments}
  
  We would like to thank Z.-F. Ji for pointing out the relationship
  between Levin-Wen entanglement entropy and the strong subadditivity
  of quantum entropy. We also thank X. Chen, X.-G. Wen, and C.-P. Sun
  for helpful discussions. YL and DLZ are supported by NSF of China
  under Grant No. 11175247, and NKBRSF of China under Grants Nos.
  2012CB922104 and 2014CB921202. BZ is supported by NSERC and CIFAR.

\end{acknowledgments}


\appendix
\section*{Appendix}

\section{Numerical results between irreducible correlations and
  topological entanglement entropy}
\label{sec:numer-results-betw}

Here we will give more numerical examples to demonstrate the
features of the irreducible many-body correlations in ground states
with topological order, and also compare
them with the corresponding numerical results of the topological
entanglement entropy. 

\begin{figure}[htbp]
  \centering 
  \includegraphics{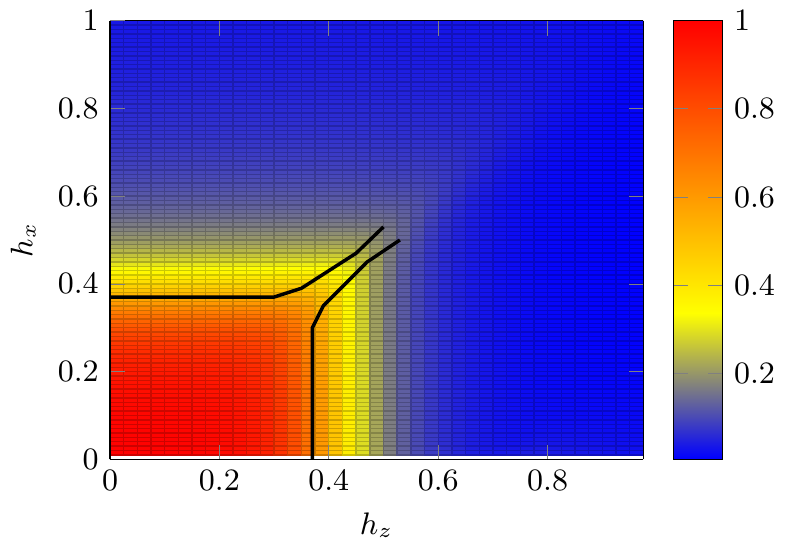}
  \caption{$C^{(6)}(\rho^{[6]})$ vs. external fields in
    $x$ and $z$ direction.}
  \label{fig:c6-hx-hz}
\end{figure}

First, we consider the toric code model on a $3\times{4}$ square
lattice with periodic condition ($24$ spins) with the external
magnetic field in the $x-z$ plane. The numerical results of
$C^{(6)}(\rho^{[6]})$ for different magnetic field is shown in
Fig.\ref{fig:c6-hx-hz}. With the increasing of the external
field, $C^{(6)}(\rho^{[6]})$ decreases from $1$ to $0$ with a
transition. The transition line is also determined by Eq. (13) and
labeled by the black lines in Fig. \ref{fig:c6-hx-hz}. The
phase diagram is similar with the previous one in Ref. \cite{TKPS2010}.

\begin{figure}[htbp]
  \centering 
  \subfloat[]{
  \includegraphics{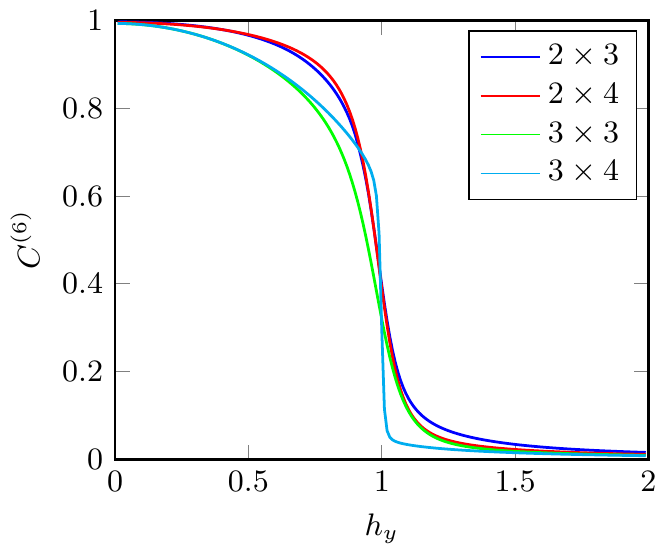}
  }
  \\ 
  \subfloat[]{
  \includegraphics{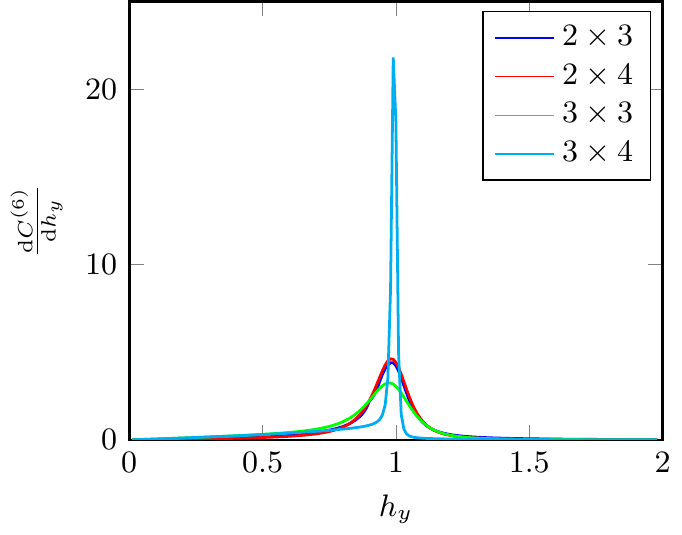}
  }
  \caption{$C^{(6)}(\rho^{[6]})$ and its derivative in the magnetic
    field $h_{y}$.}
 \label{fig:c6-hy}
\end{figure}

\begin{figure}[htbp]
  \centering 
  \subfloat[]{
    \includegraphics{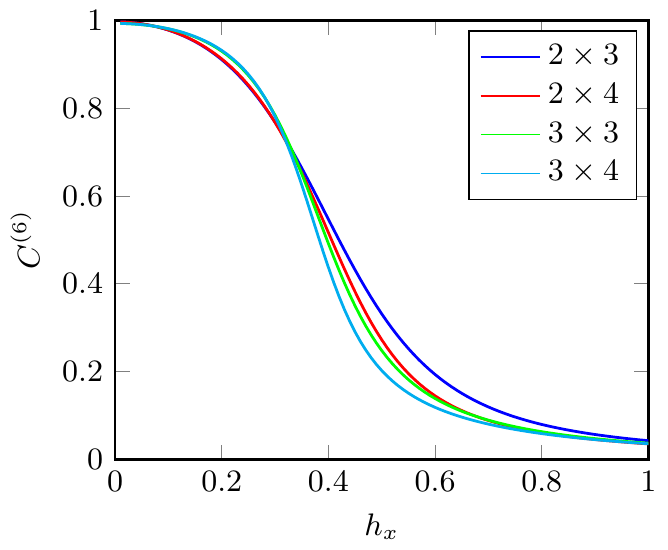}
  }
  \\
  \subfloat[]{
    \includegraphics{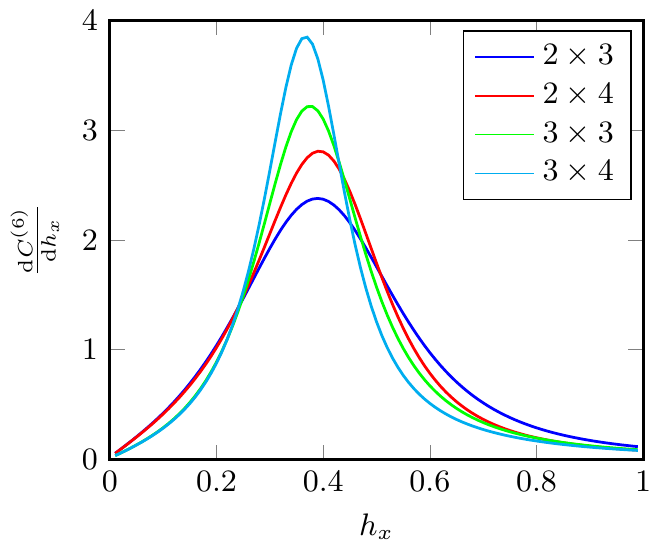}
  }
  \caption{$C^{(6)}(\rho^{[6]})$ and its derivative in the magnetic
    field $h_{x}$.}
  \label{fig:c6-hx}
\end{figure}

Second, we study how the size of the lattice affect our
results of $C^{(6)}(\rho^{[6]})$. The numerical results of
$C^{(6)}(\rho^{[6]})$ for the square lattices ($2\times{3}$,
$2\times{4}$, $3\times{3}$, and $3\times{4}$) with the magnetic field
along the $y$ direction or the $x$ direction are demonstrated in
Fig. \ref{fig:c6-hy} and Fig. \ref{fig:c6-hx}. 

In Fig.~\ref{fig:c6-hy}, there is a peak near the transition point
$h_{y}=1$ for different lattice size in the variant rate of
$C^{(6)}(\rho^{[6]})$. When the lattice size becomes $3\times{4}$, the
transition becomes very shap, which clearly identify the topological
transition.

In Fig.~\ref{fig:c6-hx}, there is a peak near the transition point in
the variant rate of $C^{(6)}(\rho^{[6]})$. When the system size
becomes larger, the peak becomes sharper, which also identify the
correct transition point.

\begin{figure}[htbp]
  \centering 
  \subfloat[]{
    \includegraphics{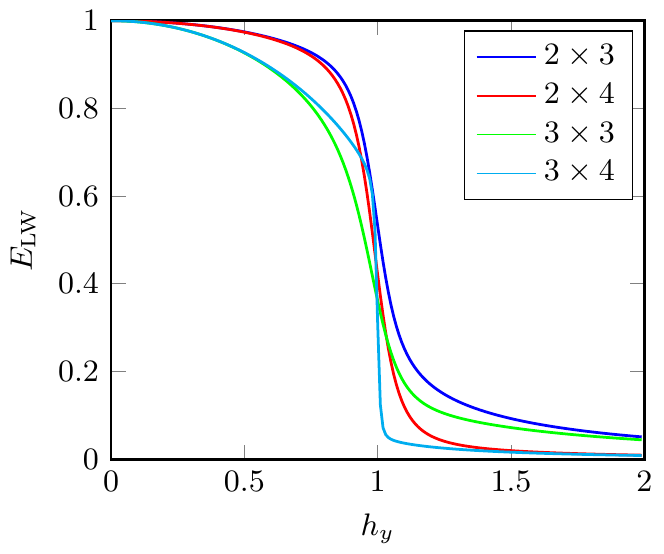}
  }
  \\ 
  \subfloat[]{
  \includegraphics{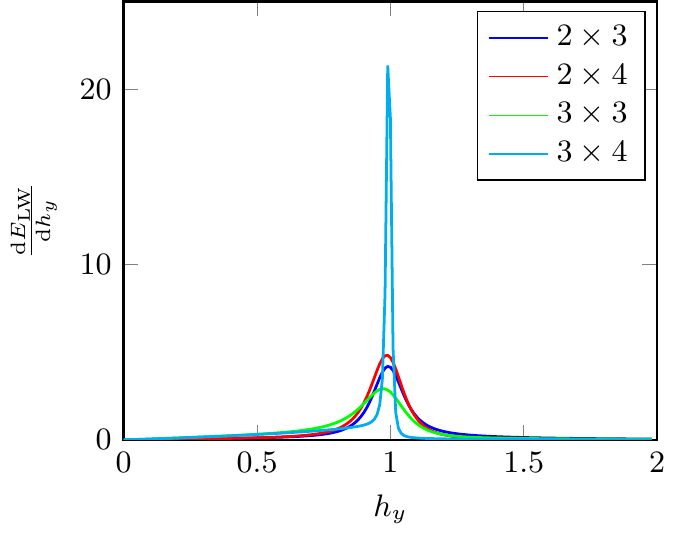}
  }
  \caption{$E_{\text{LW}}(\rho^{[6]})$ and its derivative in the magnetic
    field $h_{y}$.}
 \label{fig:elw-hy}
\end{figure}

\begin{figure}[htbp]
  \centering 
  \subfloat[]{
    \includegraphics{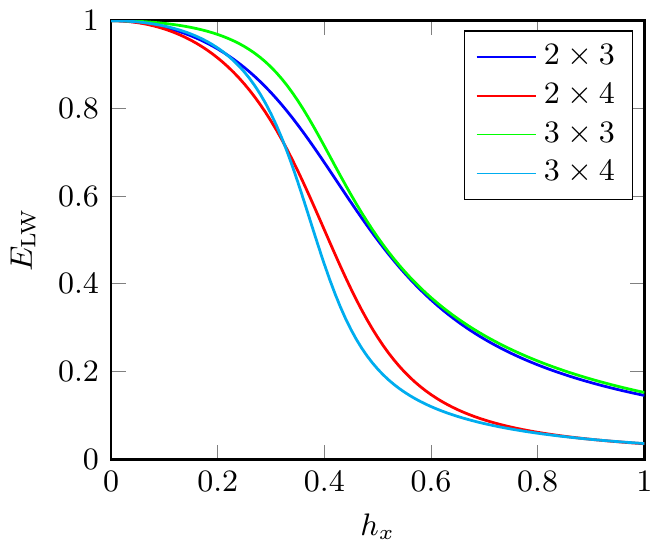}
  }
  \\
  \subfloat[]{
    \includegraphics{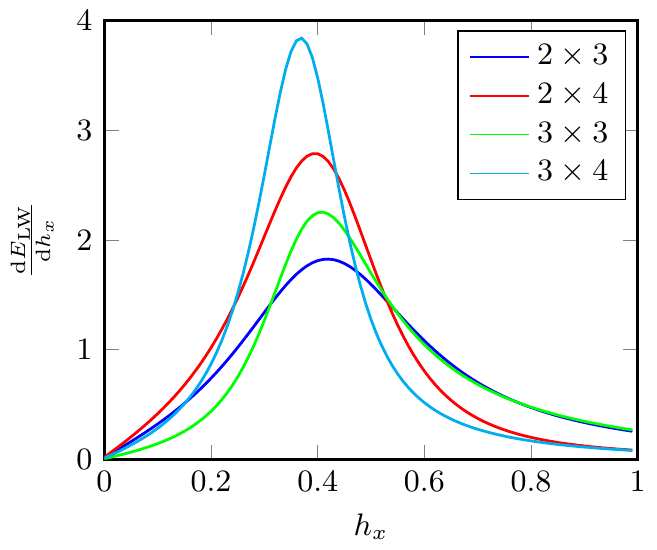}
  }
  \caption{$E_{\text{LW}}(\rho^{[6]})$ and its derivative in the magnetic
    field $h_{x}$.}
  \label{fig:elw-hx}
\end{figure}

To compare our measure with enatnglement of topological entropy, we
obtain the numerical results of $E_{\text{LW}}(\rho^{[6]})$ in the above
two cases, which are shown in Fig.~\ref{fig:elw-hy} and
Fig.~\ref{fig:elw-hx}. When the magnetic field is along the $y$
direction, we observe that $E_{\text{LW}}$ shows almost the same
behavior as $C^{(6)}$, which is consistent with the argument we
present in the article. However, when the magnetic field is along the
$x$ direction, the behaviors between $E_{\text{LW}}$ and $C^{(6)}$
show obvious differences, particularly in the smaller size of the
system. The relation between these two quantities will be studied
further in the next section.

\section{When will irreducible correlation coincide with topological
entropy}

The numerical results demonstrated above implies  that in many cases
the irreducible multiparty correlation and the topological
entanglement entropy proposed by Levin-Wen are very similar, which
motivates us to ask when will the irreducible correlations coincide
with topological entropy.

Let $\rho_{ABC}$ be a tripartite state of systems $A,B,C$. Define the
irreducible correlation of $\rho_{ABC}$ (given $\rho_{AB}, \rho_{BC}$)
as
\begin{eqnarray*}
  \label{eq:ic}
  E_{\text{IC}}(\rho_{ABC}) & = & \max \{ S(\sigma_{ABC}) \mid \sigma_{AB} =
  \rho_{AB}, \sigma_{BC} = \rho_{BC} \} \\
  & & - S(\rho_{ABC}).
\end{eqnarray*}
The Levin-Wen topological entropy of $\rho$ is
\begin{equation*}
  \label{eq:lw}
  E_{\text{LW}}(\rho_{ABC}) = S(\rho_{AB}) + S(\rho_{BC}) - S(\rho_{B}) -
  S(\rho_{ABC}).
\end{equation*}

First, we prove that $E_{\text{IC}}(\rho_{ABC})$ is upper bounded by
$E_{\text{LW}}(\rho_{ABC})$. Consider the strong subadditivity for the
maximum entropy state $\tilde{\rho_{ABC}}$, we have
\begin{equation*}
  S(\tilde{\rho}_{AB}) + S(\tilde{\rho}_{BC}) - S(\tilde{\rho}_{ABC}) -
  S(\tilde{\rho}_{B}) \ge 0.
\end{equation*}
This reduces to
\begin{equation*}
  S(\tilde{\rho}_{ABC}) \le S(\rho_{AB}) + S(\rho_{BC}) - S(\rho_B),
\end{equation*}
which is exactly
\begin{equation*}
  E_{\text{IC}}(\rho_{ABC}) \le E_{\text{LW}}(\rho_{ABC}).
\end{equation*}

From the above discussion we know that the two quantities coincide if
and only if the equality condition for strong subadditivity holds for
the maximum entropy state $\tilde{\rho}_{ABC}$.

Following Theorem 3 from Ref.~\cite{HJPW04}, we know that this happens
when the state is a quantum Markov state. In particular, the Eq.~(11)
of that reference states that this happens when
\begin{equation*}
  \tilde{\rho}_{ABC} = (\text{id} \otimes \hat{R}) \rho_{AB}.
\end{equation*}
More explicitly, this means that
\begin{equation*}
  \tilde{\rho}_{ABC} = (I_A\otimes\rho^{1/2}_{BC}) \bigl[
  (I_A\otimes\rho^{-1/2}_B) \rho_{AB} (I_A\otimes\rho^{-1/2}_B)
  \otimes I_C \bigr] (I_A\otimes\rho^{1/2}_{BC}).
\end{equation*}
One can verify that the right hand side is indeed a quantum state
whose reduced density matrix on $BC$ is exactly $\rho_{BC}$. But it is
not always the case that the reduced density matrix on $AB$ is also
$\rho_{AB}$. 
In fact, the following condition is
necessary and sufficient for the equality of $E_{\text{IC}}$ and
$E_{\text{LW}}$:

\begin{equation*}
 \tilde{\rho}_{AB} = \rho_{AB}.
\end{equation*}

Notice that
\begin{eqnarray*}
  \label{eq:1}
  C^{(3)}(\rho_{ABC}) & = & \max \{ S(\sigma_{ABC}) \mid \sigma_{AB} =
  \rho_{AB}, \sigma_{BC} = \rho_{BC}, \\
  & &  \sigma_{AC} = \rho_{AC} \} - S(\rho_{ABC}).  
\end{eqnarray*}
Therefore we have
\begin{equation*}
  \label{eq:2}
  C^{(3)}(\rho_{ABC}) \le E_{\text{IC}}(\rho_{ABC}) \le E_{\text{LW}}(\rho_{ABC}).
\end{equation*}
The condition for $C^{(3)}(\rho_{ABC})=E_{\text{LW}}(\rho_{ABC})$ is
\begin{equation*}
  \tilde{\rho}_{AB}=\rho_{AB}, \quad \tilde{\rho}_{AC}=\rho_{AC}.
\end{equation*}

In Corollary 7 of Ref.~\cite{HJPW04}, it is mentioned that a necessary
condition is that $\rho_{AC}$ is separable, i.e.,
$\rho_{AC}=\rho_A\otimes\rho_C$. But generally the product form may be
neither sufficient nor necessary.

Notice that the above discussion is valid for all the three-party
states of finite dimension. It is certainly correct to the
configuration in Fig. 5 for the Levin-Wen topological entropy.

\bibliography{arxiv}

\end{document}